\documentclass[twocolumn,A4paper,preprintnumbers,amsmath,superscriptaddress,amssymb]{revtex4-1}
\usepackage{epsfig}
\usepackage{amssymb}
\usepackage{graphicx}
\usepackage{dcolumn}
\usepackage{bm}

\begin{document}

\title{High spin polarization of the anomalous Hall current in Co-based Heusler compounds}

\author{Jen-Chuan Tung}
\affiliation{Graduate Institute of Applied Physics, National Chengchi University, Taipei 11605, Taiwan}
\author{Guang-Yu Guo}\email{gyguo@phys.ntu.edu.tw}
\affiliation{Graduate Institute of Applied Physics, National Chengchi University, Taipei 11605, Taiwan}
\affiliation{Department of Physics and Center for Theoretical Sciences, National Taiwan University, Taipei 10617, Taiwan}



\begin{abstract}
Based on first principles density functional calculations of the intrinsic anomalous and spin Hall conductivities, 
we predict that the charge Hall current in Co-based full Heusler compounds Co$_2$XZ (X = Cr and Mn; Z = Al, Si, Ga, Ge, In and Sn) 
except Co$_2$CrGa would be almost fully spin-polarized even although Co$_2$MnAl, Co$_2$MnGa, Co$_2$MnIn and Co$_2$MnSn do not 
have a half-metallic band structure. Furthermore, the ratio of the associated spin current to the charge Hall current
is slightly larger than 1.0. This suggests that these Co-based Heusler compounds, especially Co$_2$MnAl, Co$_2$MnGa and
Co$_2$MnIn which are found to have large anomalous and spin Hall conductivities, might be called anomalous Hall half-metals
and could have valuable applications in spintronics 
such as spin valves as well as magnetoresistive and spin-torque driven nanodevices. These interesting findings
are discussed in terms of the calculated electronic band structures, magnetic moments and also anomalous 
and spin Hall conductivities as a function of the Fermi level. 

\end{abstract}






\maketitle

\section{Introduction}

Spin electronics (or spintronics)\cite{Wol01} has recently become an emergent field 
because of the exciting promise of such spin-transport devices as 
magnetic field sensors for reading magnetically stored 
information based on giant magnetoresistance (GMR)\cite{Gru86,Bai88}, spin valves based tunneling 
magnetoresistance (TMR)\cite{Moo95,Parkin04}, and 
spin-torque switching-based magnetic nanodevices such as magnetic random access memories 
(MRAM)\cite{Slo96,Mye99}.
The materials which can provide a highly spin-polarized currrent, are a key ingredient for spintronics.
In this context, half-metallic ferromagnets\cite{Groot}, which are characterized by the
coexistence of metallic behavior for one spin channel and insulating behavior for the other, 
are particularly attracting.
Their electronic density of states at the Fermi level is completely spin polarized, and thus 
they could in principle offer a fully spin-polarized current. 
Therefore, since the first prediction of half-metallicity for the half Heusler compound NiMnSb in 1983\cite{Groot},
intensive research on half-metallic materials has been 
carried\cite{Sch86,Brown00,Galanakis02,Jen03,Wan06,Kubler07,Kan07}. Indeed, a large number of
materials have been predicted to be half-metallic and the half-metallicity of some of these
such as CrO$_2$ with Curie temperature $T_C = 392$ K\cite{Ji01}, has also been verified experimentally. 

Most Co-based full Heusler compounds in the cubic L2$_1$ structure are 
ferromagnetic with a high Curie temperature and a significant magnetic moment.\cite{Brown00} 
In particular, Co$_2$MnSi has the Curie temperature as high as 985 K 
and a large magnetic moment of 4.96 $\mu_B$. 
Furthermore, many of them were predicted to be half-metallic\cite{Galanakis02,Kubler07,Kan07} 
and hence are of particular interest for spintronics. 
Therefore, the electronic band structure and magnetic properties of the full Heusler compounds 
have been intensively investigated both theoretically and 
experimentally in recent years\cite{Brown00,Galanakis02,Kubler07,Kan07}.
For example, the total magnetic moments of these materials were found to follow the Slater-Pauling type
behavior and the mechanism was explained in terms of the calculated electronic structures\cite{Galanakis02}. 
The Curie temperatures of Co-based Heulser compounds were also determined {\it ab initio} and
the trends were related to the electronic structures\cite{Kubler07}.

In this paper, we study anomalous Hall effect (AHE) and spin-polarization of Hall current
in the Co-based full Heusler compounds Co$_2$XZ (X = Cr and Mn; Z = Al, Si, Ga, Ge, In and Sn) 
by {\it ab initio} calculations of intrinsic anomalous and spin Hall conductivities.
Anomalous Hall effect, discovered in 1881 by Hall\cite{Hal81}, is an archetypal 
spin-related transport phenomenon and hence has received renewed attention in recent years\cite{Nagaosa10}.
Indeed, many {\it ab initio} studies on the AHE in elemental ferromagnets\cite{Yao04,Rom09,Fuh11,Tung12} 
and intermetallic compounds\cite{zen06,He12} have recently been reported. 
However, first principles investigations into the AHE in half-metallic ferromagnets,
which is interesting on its own account, have been very few\cite{Kubler12}. 
Therefore, a principal purpose of this work is to understand the AHE in the Co-based
full Heusler compounds especially those of half-metallic ones. The results may also help 
experimentally search for the Heusler compounds with large AHE for applications, e.g., in magnetization sensors\cite{Vidal11}.

The intrinsic AHE is caused by the opposite anomalous velocities experienced by 
spin-up and spin-down electrons when they move through the relativistic energy bands in solids
under the influence of the electric field.\cite{Nagaosa10} In ferromagnets, where an unbalance of spin-up and 
spin-down electrons exists, these opposite transverse currents would give rise to
the spin-polarized charge current (i.e., anomalous Hall current). In nonmagnetic materials where
the numbers of the spin-up and spin-down electrons are equal, the same process would result in
a pure spin current, and this is known as the intrinsic spin Hall effect (SHE).\cite{Mur03}
Interestingly, the pure spin current is dissipationless\cite{Mur03} and is thus
important for the development of low energy-consumption nanoscale spintronic devices\cite{Liu12}.
We note that high spin-polarization ($P$) of the charge current ($I_C$) from the electrode is
essential for large GMR and TMR. However, since the current-induced magnetization switching 
results from the transfer of spin angular momentum from the current carriers to the magnet\cite{Slo96}, 
large spin current ($I_S$) would be needed for the operation of the spin-torque 
switching-based nanodevices\cite{Slo96,Mye99}, i.e., a large ratio of spin current 
to charge current [$\eta = |(2e/\hbar) I_S/I_C|$], would be crucial.   
For ordinary charge currents, this ratio $\eta$ varies from 0.0 (spin unpolarized current) 
to 1.0 (fully spin polarized current). Interestingly, $\eta$ can be larger than 1.0 for the charge Hall 
currents and is $\infty$ for pure spin current. Excitingly, spin-torque switching of ferromagnets
driven by intense pure spin current from the large spin Hall effect in tantalum has been recently reported\cite{Liu12}. 
Therefore, because of this and also the topological nature of the intrinsic AHE\cite{Nagaosa10},
it might be advantageous to use the Hall current from ferromagnets for magnetoelectronic devices, 
rather than the longitudinal current. Another purpose of this work is therefore to 
investigate the nature and spin-polarization of the Hall current in the
Co-based Heusler compounds, the knowledge of which is required for possible spintronic applications of the Hall current. 


\section{Theory and computational method}

The intrinsic anomalous and spin Hall conductivities of a solid can be evaluated by using the Kubo 
formalism\cite{Yao04,Fuh11,Tung12,Guo05}. The intrinsic Hall effect comes from the static limit 
($\omega =0$) of the off-diagonal elements of the optical conductivity\cite{Guo05}. 
Here we first calculate the imaginary part of the off-diagonal elements of the optical conductivity.
Then we obtain the real part of the off-diagonal elements from the corresponding imaginary part 
by a Kramers-Kroning transformation. The intrinsic Hall conductivity $\sigma^{H}_{xy}$ 
is the static limit of the off-diagonal element of the optical conductivity $\sigma^{(1)}_{xy}(\omega=0)$ 
(see Ref. \cite{Tung12} for more details). We notice that the anomalous Hall conductivity (AHC) of 
bcc Fe\cite{Yao04} and the spin Hall conductivity (SHC) of fcc Pt\cite{Guo08} calculated 
in this way are in good agreement with that calculated directly by accounting for the Berry 
phase correction to the group velocity. 

Since all the intrinsic Hall effects are caused by the spin-orbit coupling (SOC), first-principles calculations 
must be based on a relativistic band theory. We calculate the relativistic band structure of the Co-based Heusler 
compounds (Co$_2$XZ) considered here using the highly accurate full-potential linearized augmented
plane wave (FLAPW) method, as implemented in the WIEN2K code\cite{wien2k02}. 
The self-consistent electronic structure calculations are based on the density 
functional theory (DFT) with the generalized gradient approximation (GGA) 
for the exchange correlation potential\cite{Perdew96}. We consider only the fully ordered cubic 
Heusler compounds structure (L2$_1$) and use the experimental lattice constants for all the  
considered Co$_2$XZ Heusler compounds except Co$_2$CrSi, Co$_2$CrGe and Co$_2$MnIn, as listed in Table 1. 
Since the lattice constants of Co$_2$CrSi, Co$_2$CrGe and Co$_2$MnIn have not been reported, 
we have determined their lattice constants theoretically\cite{Tung12b}. 
The SOC is included using the second variation technique \cite{wien2k02} with the magnetization 
along the $c$-axis in all the present calculations. The wave function, charge 
density, and potential were expanded in terms of the spherical harmonics inside the muffin-tin spheres 
and the cutoff angular moment ($L_{max}$) used is 10, 6 and 6, respectively. The wave function outside the 
muffin-tin spheres was expanded in terms of the augmented plane waves (APW) and a large number of augmented plane waves 
(about 70 APWs per atom, i. e., the maximum size of the crystal momentum $K_{max}=8/R_{mt}$) were included 
in the present calculations. The improved tetrahedron method is used for the Brillouin-zone
integration\cite{Blochl94}. To obtain accurate ground state charge density as well as spin and orbital 
magnetic moments, a fine 36$\times$36$\times$36 grid with 1240 $k$-points in the irreducible wedge of the first Brillouin zone was used.

\section{Magnetic moments and half-metallicity}

\begin{table*}
\caption{Calculated total spin magnetic moment ($m_s^{tot}$) ($\mu_B$/f.u.), 
atomic spin magnetic moment ($m_s$) and orbital magnetic moment (($m_o$) ($\mu_B$/atom) as well as spin-decomposed
density of states at the Fermi level [$N^{\uparrow}(E_F)$, $N^{\downarrow}(E_F)$] (states/eV/f.u.) of all the considered
Heusler compounds Co$_2$XZ, together with the lattice constants $a$ (\AA) used and the number of the valence electrons ($n_v$) 
per formula unit (f.u.).  For comparison, the available experimental magnetic moments 
are also listed. The orbital magnetic moment for the non-transition metal atoms ($m_{o}^{Z}$) is negligible, i.e., 
being less than 0.0001 $\mu_B$/atom, and hence not listed here. }
\begin{center}
\begin{tabular}{ccccccccccc}\hline\hline
Co$_2$XZ   &      $a$      &$n_v$&$m_{s}^{tot}$&$m_{s}^{Co}$ &$m_{s}^X$ &$m_{s}^Z$ &$m_{o}^{Co}$ &$m_{o}^{X}$ &$N^{\uparrow}(E_F)$& $N^{\downarrow}(E_F)$\\ \hline
Co$_2$CrAl &5.727\footnotemark[1]&27&   2.998            & 0.804       & 1.533    &  -0.061  &   0.016     &  0.007     &    4.160        &    0.000       \\
           &               &&1.65\footnotemark[2]&             &          &          &             &            &                 &                  \\
Co$_2$CrSi &5.630\footnotemark[2]&28&  3.997      & 1.013       & 2.015    &  -0.045  &   0.025     & -0.008     &    3.064        &  0.000   \\
Co$_2$CrGa &5.805\footnotemark[1]&27&  3.051      & 0.758       & 1.651    &  -0.053  &   0.016     &  0.006     &    2.928        &  0.215 \\                
Co$_2$CrGe &5.740\footnotemark[2]&28&  3.997      & 0.950       & 2.129    &  -0.033  &   0.025     & -0.013     &    2.820        &  0.000 \\
Co$_2$MnAl&5.755\footnotemark[3]&28& 4.037     & 0.760       & 2.730    &  -0.079  &   0.011     &  0.019     &    1.183        &  0.229\\
          &               & &4.07\footnotemark[3]&         &          &          &             &            &     &   \\
Co$_2$MnSi &5.654\footnotemark[4]&29&   4.997     & 1.059       & 2.995    &  -0.052  &   0.027     &  0.015     &    1.254        &  0.000  \\
           &               & &4.96\footnotemark[4]&            &          &          &             &            &                 &                \\
Co$_2$MnGa &5.770\footnotemark[4]&28&   4.128     & 0.754       & 2.794    &  -0.062  &   0.009     &  0.021     &    1.832        &  0.369 \\
           &               & &3.72\footnotemark[4]&            &          &          &             &            &                 &        \\                
Co$_2$MnGe &5.743\footnotemark[4]&29&   4.999     & 0.999       & 3.091    &  -0.041  &   0.030     &  0.019     &    1.290        &  0.000 \\
           &               &&4.84\footnotemark[4]&            &          &          &             &            &                 &        \\
Co$_2$MnIn &5.990\footnotemark[2]&28&   4.460     & 0.812       & 3.034    &  -0.056  &   0.017     &  0.023     &    2.321        & 1.235  \\
Co$_2$MnSn &6.000\footnotemark[4]&29&   5.033     & 0.974       & 3.241    &  -0.046  &   0.034     &  0.021     &    1.215        & 0.175  \\
           &               &&4.78\footnotemark[4]&            &          &          &             &            &        &               \\ \hline \hline
\end{tabular}
\end{center}
\footnotetext[1]{Experimental data (Ref. \cite{Bus83}).}
\footnotetext[2]{GGA calculations (Ref. \cite{Tung12b}).}
\footnotetext[3]{Experimental data (Ref. \cite{Ume08}).}
\footnotetext[4]{Experimental data (Ref. \cite{Brown00}).}
\end{table*}

\begin{figure}
\begin{center}
\includegraphics[width=7cm]{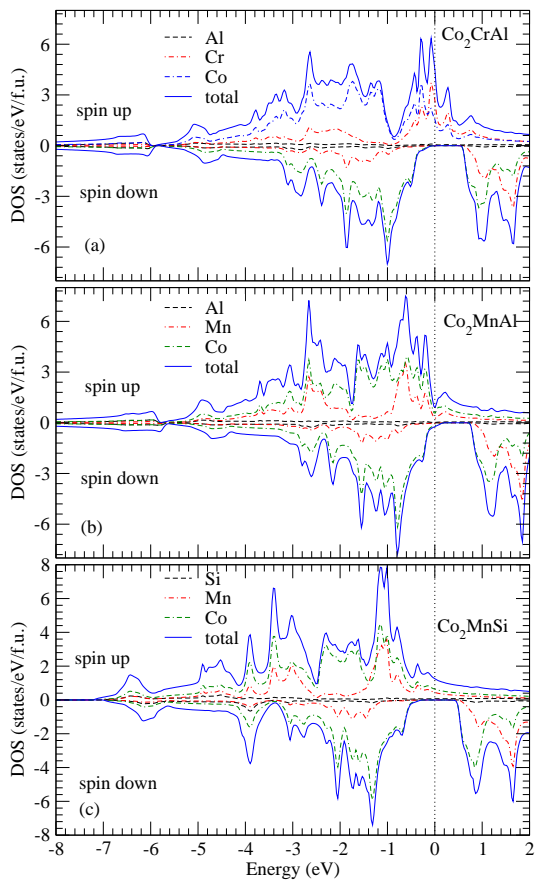}\
\caption{(color online) Total and site decomposed densities of states (DOSs) for 
(a) Co$_2$CrAl, (b) Co$_2$MnAl and (c) Co$_2$MnSi. The Fermi level is set to zero.}
\end{center}
\end{figure}

Let us first examine the calculated magnetic properties and band structures near the Fermi level 
of the considered Co-based Heusler compounds. Since the electronic structure and magnetism in
the full Heusler compounds have been extensively studied (see, e.g., Refs. \cite{Galanakis02},
\cite{Kubler07}, \cite{Kan07} and references therein), here we focus on only the salient features
which may be related to the anomalous and spin Hall effects as well as spin polarization of the
Hall current to be presented in the next two sections. The calculated total spin magnetic moment, 
local spin and orbital magnetic moments as well as spin-decomposed densities of states (DOSs)
at the Fermi level ($E_F$) of all the considered Co-based Heusler compounds are listed in Table 1, 
together with  the available experimental total spin magnetic moments for comparison.  
The site-decomposed DOSs for the three representative Heusler compounds 
Co$_2$CrAl, Co$_2$MnAl and Co$_2$MnSi are dislayed in Fig. 1.

First of all, Table 1 shows that Co$_2$CrZ (Z = Al, Si and Ge) and Co$_2$MnZ (Z = Si and Ge) are half-metallic,
since the spin-down DOS at the Fermi level [$N^{\downarrow}(E_F)$] for these 
compounds is zero. Therefore, their calculated total spin magnetic moments are almost an integer, 
as all the half-metallic ferromagnets should be.
The tiny deviation of the total spin magnetic moment from an integer for these
compounds is due to the inclusion of the SOC in the present calculations. 
Note that, in principle, including the SOC could induce a nonvanishing $N^{\downarrow}(E_F)$ 
in the spin-down band gap of half-metals and hence could destroy the half-metallicity. 
Nevertheless, the $N^{\downarrow}(E_F)$ was found to be very small in Heusler compound NiMnSb 
although it could be large in heavy element compounds such as MnBi\cite{Mav04}.
Indeed, the $N^{\downarrow}(E_F)$ in the above mentioned Co$_2$XZ compounds obtained with the SOC is
negligible (see Table 1). 
Co$_2$CrGa, Co$_2$MnAl,  Co$_2$MnGa, Co$_2$MnSn and Co$_2$MnIn are not half-metallic (see Table 1 and Fig. 1).
Nevertheless, in Co$_2$CrGa, Co$_2$MnAl and Co$_2$MnSn, because the $E_F$ is located only 
slightly below the top of the spin-down valence band (Fig. 1), the total spin magnetic moment
deviates only slightly from an integer (see Table 1 and Fig. 1). 

Secondly, we find that the local spin magnetic moments on the 3$d$ transition metal sites, namely,
Co and X (X = Cr and Mn), are large and coupled ferromagnetically (see Table 1).
The Co atoms have a spin magnetic moment ranging 0.7 to 1.0 $\mu_B$, and the spin magnetic moment
of the Mn (Cr) atoms is around 3.0 (2.0) $\mu_B$. In contrast, the local spin magnetic moments on
the non-transition metal atoms Z (Z = Al, Si. Ga, Ge, In and Sn) are small and aligned antiparallel to
that of the Co and X atoms. 
Not surprisingly, the orbital magnetic moments on the transition metal atoms (Co, Mn and Cr)
are rather small, being about 2 order of magnitude smaller than the spin magnetic moments,
because of the weakness of the SOC in these 3$d$ transition metal compounds. 
All the non-transition metal (Z) atoms have a practically zero orbital magnetic moment (i.e., being
$\le$ 0.0001 $\mu$$_B$/atom) and thus are not listed in Table 1. The calculated total spin magnetic
moments are in good agreement with previous theoretical calculations\cite{Galanakis02,Kubler07} 
and also the experimental results\cite{Brown00}.

Thirdly, we notice that the calculated total spin magnetic moments for all the Heusler compounds 
more or less follow the so-called $n_v$-24 rule ($n_v$ is the number of valence electrons), 
as has already been reported in Ref. \cite{Galanakis02}. 
For 3$d$ transition metals and their binary compounds, the total spin magnetic moment ($m_s^{tot}$) shows 
the well-known Slater-Pauling 
behavior\cite{Kit68}. This $m_s^{tot} = n_v$-24 rule is essentially a generalized Slater-Pauling
rule for the full-Heusler compounds. The reason is that in these compounds, the number of occupied
spin-down states ($n^{\downarrow}$) is found to remain fixed at 12, at least when they are in the
half-metallic state.\cite{Galanakis02} 
Therefore, $m_s^{tot} = n_v-2n^{\downarrow} = n_v-24$. For example, the $n_v$ for Co$_2$CrAl and Co$_2$CrGa is 27, 
and the calculated $m_s^{tot}$ are 3  $\mu_B$. Similarly, the total spin magnetic moment is 
4 $\mu_B$ for Co$_2$MnAl, and 5 $\mu_B$ for Co$_2$MnSi, Co$_2$MnGe, Co$_2$MnSn, also 
following this $n_v-24$ rule. The obvious exceptions are Co$_2$MnGa and Co$_2$MnIn because they
deviate strongly from the half-metallicity (Table 1). 

Displayed in Fig. 1 are the total and site decomposed DOSs of three selected Heusler compounds 
Co$_2$CrAl (a), Co$_2$MnAl (b) and Co$_2$MnSi (c). Fig. 1 shows clearly that Co$_2$CrAl and Co$_2$MnSi are
half-metallic with the $E_F$ falling in the spin-down insulating gap, while Co$_2$MnAl is
a normal ferromagnetic metal with the $E_F$ being located just below the top of the
spin-down valence band. The DOS spectra for Co$_2$MnAl and Co$_2$MnSi are in general similar except 
the location of the Fermi level. As one goes from Co$_2$MnSi to Co$_2$MnAl, the $n_v$ 
decreases by one and hence the $E_F$ is lowered from the middle of the spin-down insulating gap to
that just below the top of the spin-down valence band. Similar situations occur for Co$_2$MnGa and Co$_2$MnGe, 
and thus their DOS spectra are not shown here. However, pronounced differences in the band structure 
between Co$_2$CrAl and Co$_2$MnAl exist. These differences are mainly caused by the different exchange splittings
of the Cr and Mn 3$d$ bands. The Mn atoms have a larger spin moment of $\sim$ 3.0 $\mu_B$ and hence
a larger 3$d$ band exchange splitting, whilst the Cr atoms have a smaller spin moment of $\sim$ 2.0 $\mu_B$ and hence
a smaller exchange splitting (see Table 1 and Fig. 1). Consequently, the Fermi level is located at the center
of the spin-up 3$d$ band in Co$_2$CrAl (Fig. 1a) whilst the 
spin-up 3$d$ band in Co$_2$MnAl is mostly below the Fermi level (Fig. 1b). 
In short, the Co-based Heusler compounds considered here can be divided into two families, namely,
Co$_2$CrZ and Co$_2$MnZ. Within each family, the electronic structure and other physical properties for
one member can be approximately obtained from another member by appropriately shifting the Fermi level. 
One exception is the pair of Co$_2$MnSn and Co$_2$MnIn.

\section{Anomalous and spin Hall conductivities}

A dense $k$ point mesh would be needed for obtaining accurate anomalous and spin Hall 
conductivities\cite{Yao04,Guo05,Fuh11}. Therefore, we use several fine $k$-point meshes with the 
finest $k$-point mesh being 58$\times$58$\times$58. To obtain the theoretical anomalous and spin 
Hall conductivities, we first calculate the AHC and SHC 
as a function of the number ($N_k$) of $k$-points in the first Brillouin zone. The calculated 
anomalous ($\sigma^{A}_{xy}$) and spin ($\sigma^{S}_{xy}$) Hall conductivities 
versus the inverse of the $N_k$ are then plotted and fitted to a first-order polynomial to 
get the converged theoretical $\sigma^{A}_{xy}$ and $\sigma^{S}_{xy}$ (i.e. the extrapolated value 
at $N_k = \infty$) (see Refs. \cite{Fuh11} and \cite{Tung12}). The theoretical $\sigma^{A}_{xy}$ and
$\sigma^{S}_{xy}$ determined this way are listed in Table 2. 

\begin{table*}
\caption{Calculated anomalous [$\sigma_{xy}^A$ (S/cm)] and spin [$\sigma_{xy}^S$ ($\hbar$S/e cm)] Hall conductivities
as well as
spin-decomposed Hall conductivities ($\sigma^{H\uparrow}_{xy}$, $\sigma^{H\downarrow}_{xy}$) (S/cm), 
Hall current spin polarization $P^H$ ($\%$), spin polarization of the electronic states at the 
Fermi level $P^{D}$ ($\%$) and the ratio of spin current to charge current $\eta$. }
\begin{center}
\begin{tabular}{cccccccc}\hline\hline

Co$_2$XZ  &$\sigma^{A}_{xy}$&$\sigma^{S}_{xy}$&$\sigma^{H\uparrow}_{ij}$&$\sigma^{H\downarrow}_{ij}$& $P^H$ & $P^{D}$ & $\eta$ \\ \hline
Co$_2$CrAl& 241 (438\footnotemark[1]) &  137    &    258     &    -17.0   &  114  &  100 (100\footnotemark[1]) & 1.14 \\
          & 125\footnotemark[2] &               &            &            &       &       & \\
Co$_2$CrSi& 175           &  101          &    189     &    -13.8   &  115  &  100  & 1.15  \\
Co$_2$CrGa& 327           &  106          &    269     &    -58.0   &   64  &   86  & 0.65   \\
Co$_2$CrGe& 234           &  133          &    250     &    -16.6   &  114  &  100  & 1.14   \\
Co$_2$MnAl& 1265 (1800\footnotemark[1])&  655   &    1288    &    -23.1   &  104  &   68 (75\footnotemark[1]) & 1.04 \\
          &1500$\sim$2000\footnotemark[3] &     &            &            &       &  & \\
Co$_2$MnSi& 193 (228\footnotemark[1])  & 110    &    206     &    -13.4   &  114  &  100 (100\footnotemark[1]) & 1.14   \\
Co$_2$MnGa& 1417          & 733           &    1441    &    -24.5   &  103  &   66  & 1.03 \\
Co$_2$MnGe& 228           & 134           &    248     &    -19.5   &  117  &  100  & 1.18 \\
Co$_2$MnIn& 926           & 433           &    895     &     29.9   &   93  &   31  & 0.94 \\
Co$_2$MnSn& 174 (118\footnotemark[1])  & 101    &    188     &    -13.9   &  116  &   75 (82\footnotemark[1]) & 1.16 \\ \hline\hline
\end{tabular}
\end{center}
\footnotetext[1]{Theoretical results (Ref. \cite{Kubler12}).}
\footnotetext[2]{Experimental data (Ref. \cite{Husmann06}).}
\footnotetext[3]{Experimental data (Ref. \cite{Vidal11}).}
\end{table*}
 
Table 2 shows that the calculated $\sigma_{xy}^A$ and $\sigma_{xy}^S$ are large for Co$_2$MnAl, Co$_2$MnGa and Co$_2$MnIn,
but are five to ten times smaller for the other compounds. This suggests that Co$_2$MnAl, Co$_2$MnGa and Co$_2$MnIn
may find applications for, e.g., magnetic sensors.\cite{Vidal11}
Interestingly, the calculated $\sigma^{S}_{xy}$ seems to be about half of the $\sigma^{A}_{xy}$ for
every compound considered here except Co$_2$CrGa. This may be attributed to the half-metallic 
behavior of these Heusler compounds.
In the half metallic materials, the charge current would flow only in one spin channel, and no charge current
in the other spin channel, resulting in $\sigma_{xy}^A$ being twice as large as $\sigma_{xy}^S$.
We will further discuss this point in the next section.

\begin{figure}
\begin{center}
\includegraphics[width=8cm]{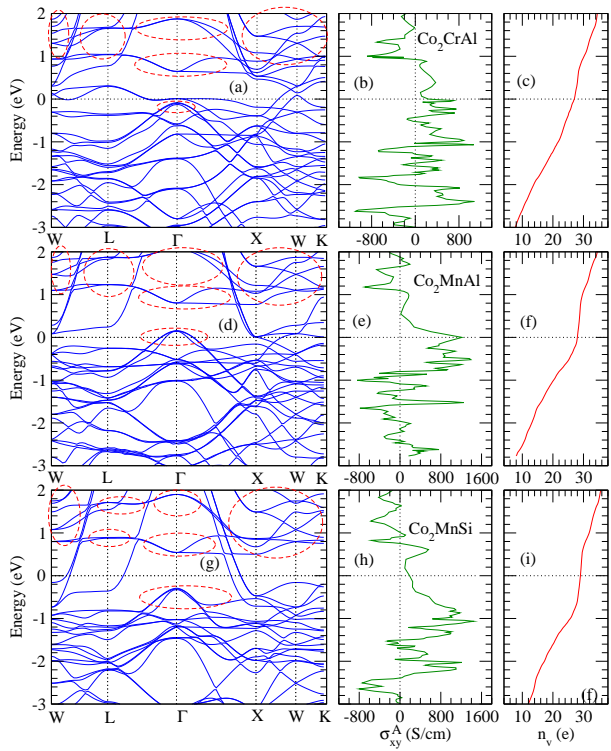}\\
\caption{(color online) Band structure (left panels), anomalous Hall conductivity ($\sigma^A_{ij}$)
(middle panels) and number of valence electrons per formula ($n_v$) (right panels)
for  Co$_2$CrAl (upper panels), Co$_2$MnAl (middle panels) and Co$_2$MnSi (bottom panels). 
The Fermi energy is shifted to zero. The red dashed ellipses in the left panels mark the
predominantly spin-down energy bands.}
\end{center}
\end{figure}

We notice that the difference in the AHC between Co$_2$MnAl (Co$_2$MnGa) and Co$_2$MnSi (Co$_2$MnGe) 
is rather dramatic while the number of valence electrons in Co$_2$MnAl (Co$_2$MnGa) differs from that 
in Co$_2$MnSi  (Co$_2$MnGe) only by one (see Table 1). In particular, the AHC in Co$_2$MnAl (Co$_2$MnGa)
is about six times larger than that in Co$_2$MnSi (Co$_2$MnGe). This is somewhat surprising because
one would expect a half-metallic metal with a larger magnetization to have a larger AHC than
a normal ferromagnet with a smaller magnetization.  
To better understand this, we display the relativistic band structure as well as the AHC and $n_v$ 
as a function of the $E_F$ for three selected compounds Co$_2$CrAl, Co$_2$MnAl 
and Co$_2$MnSi in Fig. 2. Fig. 2(d) and (g) suggest that the band structures of Co$_2$MnAl
and Co$_2$MnSi are rather similar. The key difference is the location of the $E_F$ due to the difference
in the $n_v$ in these compounds. In other words, the band structure and other physical
properties of Co$_2$MnSi may approximately be obtained from that of Co$_2$MnAl by raizing the $E_F$ by
about 0.5 eV due to one extra $p$ valence electron. Interestingly, there is a pronounced peak sitting at 
the $E_F$ in the $\sigma_{xy}^A$ spectrum of Co$_2$MnAl. This feature may be attributed to the prominant 
contributions to the AHC from the spin-down band pocket at the $\Gamma$ point and also the narrow spin-up
bands along the Brillouin zone edges X-W and W-K (see Fig. 2, and also Fig. 5 in Ref. \cite{Kubler12}). 
These band features also apppear in the isoelectronic compound Co$_2$MnGa 
(not shown here) which has a large AHC too (Table 2). However, when the Fermi level is raized to the 
highly dispersive Co and Mn $d$-orbital hybridized spin-up band region to accommodate 
one more valence electron as one goes from Co$_2$MnAl (Co$_2$MnGa)
to Co$_2$MnSi (Co$_2$MnGe), these band features are now significantly below the $E_F$ (Fig. 2) 
with much diminished contributions to the AHC, thus resulting in the much reduced $\sigma_{xy}^A$ 
and $\sigma_{xy}^S$ in Co$_2$MnSi (Co$_2$MnGe) (Table 2). 
Interestingly, the calculated $\sigma_{xy}^A$ is larger in Co$_2$MnGa than in Co$_2$MnAl.
This difference in the $\sigma_{xy}^A$ could be attributed to the larger SOC in the Ga atoms than in the Al atoms.  
This finding prompted us to calculate the $\sigma_{xy}^A$ and $\sigma_{xy}^S$ for Co$_2$MnIn.
Unfortunately, the calculated $\sigma_{xy}^A$ in Co$_2$MnIn is even smaller than that of Co$_2$MnAl.

However, as described in the preceding section, significant differences in the band structure 
between Co$_2$MnAl and Co$_2$CrAl exist and they are not simply due to the different locations of the $E_F$ in
these compounds (Fig. 2). In particular, the top of the spin-down valence band 
which sticks out of the Fermi level in Co$_2$MnAl becomes completely submerged below the Fermi level
in Co$_2$CrAl [Fig. 2(a) and (d)]. These result in rather pronounced differences in the $\sigma_{xy}^A$
at the $E_F$ and below. For example, the large broad peak in the $\sigma_{xy}^A$ spectrum between
-0.8 and 0.0 eV in Co$_2$MnAl (Fig. 2e) becomes a much reduced narrow peak in Co$_2$CrAl (Fig. 2b),
and Co$_2$CrAl has a much smaller $\sigma_{xy}^A$ value of 241 S/cm (Table 2). Nevertheless, the
calculated $\sigma_{xy}^A$ spectra above the $E_F$ in the two compounds are rather similar,
both becoming negative at $\sim$1.0 eV.

The anomalous Hall conductivity for Co$_2$CrAl was experimentally found to follow closely the magnetization
with the ratio ($\sigma_H^1$) of the two quantities being constant over a large temperature 
range\cite{Husmann06}, suggesting the intrinsic origin. Nonetheless, the measured AHC 
conductivity for Co$_2$CrAl is $\sim$125 S/cm\cite{Husmann06}, being only about half of our theoretical 
value (Table 2). 
We notice that the measured magnetization is only 1.65 $\mu_B$/f.u.\cite{Husmann06}, being about two times 
smaller than our theoretical value of 3.0 $\mu_B$/f.u. (Table 1). Previous experimental studies and
{\it ab initio} calculations showed that the intrinsic AHC is proportional to the magnetization.\cite{zen06,Husmann06}  
Therefore, the smaller experimental AHC may be attributed to the smaller observed magnetization in Co$_2$CrAl
due to the presence of imperfections such as structural disorder.\cite{kud08} 
If we use the measured $\sigma_H^1$ and extrapolate the linear relation $\sigma_{xy}^A = \sigma_H^1\times M$
to the theoretical magnetization ($M$) value, we get an "experimental" $\sigma_{xy}^A$ value of 227 S/cm,
being quite close to the theoretical $\sigma_{xy}^A$ value of 241 S/cm. 
The AHE in Co$_2$MnAl was recently measured by
Vidal et al.\cite{Vidal11}, and the Hall resistivity $\rho_{xy}$ was found to be about 20 and 15  $\mu$$\Omega$cm 
in clean and dirty conditions, respectively. Using these experimental values together with the 
longitudinal resistivity of about 100 $\mu$$\Omega$cm, one can estimate the AHC to range from 1500 to 2000 S/cm.
We notice that our theoretical $\sigma^{A}_{xy}$ of 1264 S/cm for Co$_2$MnAl is smaller than
the estimated experimantal values\cite{Vidal11} by about 15$\sim$40 \%.
The discrepancy could be due to the fact that the Co$_2$MnAl samples used in the experiments contain mostly
the B2 phase\cite{Vidal11}, instead of the L2$_1$ structure considered here.
Note that the measured AHC values would contain both the intrinsic $\sigma_{xy}^A$
and the contributions from extrinsic skew scattering and side jump mechanisms\cite{Nagaosa10}. 
Therefore, another source of the discrepancy could come from the extrinsic AHC 
which is not addressed here. In fact, recent {\it ab initio} calculations based 
on the short-range disorder in the weak scattering limit\cite{Wei11} indicated that
the scattering-independent side jump contribution is in the order of $\sim$100 S/cm
in 3$d$ transition metal ferromagnets and L1$_0$ FePd and FePt. 
Similar values of the extrinsic AHC in L1$_0$ FePd$_{1-x}$Pt$_x$ alloys were derived
recently by comparing the measured AHC with the calculated intrinsic AHC\cite{He12,See10}.  

Our theoretical $\sigma^{A}_{xy}$ value for Co$_2$MnSi agrees rather well with the theoretical value obtained recently
by computing Berry curvatures\cite{Kubler12} (Table 2). However, for Co$_2$CrAl, Co$_2$MnAl and Co$_2$MnSn, 
our theoretical $\sigma^{A}_{xy}$ value differs significantly from that reported in Ref. \cite{Kubler12} by
about 30$\sim$100 \% (see Table 2). We notice that in Ref. \cite{Kubler12}, only 2000 $k$-points 
in the Brillouin zone were used and the convergence of the $\sigma^{A}_{xy}$ was reported to be about 20 $\%$. 
Also, here we used
the accurate FLAPW method whilst, in contrast, the atomic spherical wave method with the atomic
sphere approximation (ASA) was used in Ref. \cite{Kubler12}. Since the calculated anomalous Hall conductivity 
is sensitive to both the number of $k$ points in the Brillouin zone used and the details of the energy 
bands in the vicinity of the Fermi level\cite{Yao04,Guo05,Fuh11,Tung12}, the discrepancies in
the calculated $\sigma^{A}_{xy}$ between the two theoretical studies may perhaps be due to 
the fewer $k$ points in the Brillouin zone and the ASA used in Ref. \cite{Kubler12}.


  
\section{Hall current spin polarization}
The anomalous Hall effect has recently received intensively renewed interests mainly because of its close 
connection with spin-transport phenomena.\cite{Nagaosa10} 
Indeed, it could be advantageous to use the Hall current from a ferromagnet as a spin-polarized 
current source, instead of the longitudinal current, as mentioned before. 
Therefore, it would be interesting to know the spin polarization of the Hall current. 
The spin polarization $P^H$ of the Hall current may be written as
\begin{equation}
P^H =\frac{ \sigma^{H\uparrow}_{xy}-\sigma^{H\downarrow}_{xy} }{\sigma^{H\uparrow}_{xy}+\sigma^{H\downarrow}_{xy} }
\end{equation}
where $\sigma^{H\uparrow}_{xy}$ and $\sigma^{H\downarrow}_{xy} $ are the spin-up and spin-down Hall conductivities, 
respectively. The $\sigma^{H\uparrow}_{xy}$ and $\sigma^{H\downarrow}_{xy}$ can be obtained from the calculated 
AHC and SHC via the relations
\begin{eqnarray}
\sigma^{A}_{xy} = \sigma^{H\uparrow}_{xy}+\sigma^{H\downarrow}_{xy} \\
2\frac{e}{\hbar}\sigma^{S}_{xy} = \sigma^{H\uparrow}_{xy}-\sigma^{H\downarrow}_{xy}.
\end{eqnarray}
Note that, the absolute value of $P^H$ can be greater than 1.0 because the spin-decomposed Hall currents 
can go either right (positive) 
or left (negative). In the nonmagnetic materials, the charge Hall current is zero, and hence,
$\sigma^{H\uparrow}_{xy}$ $=$ $-\sigma^{H\downarrow}_{xy} $ results in $P^H$$=\infty$. 
Also of interest is the ratio of the spin current to charge current $\eta$ which can be written 
as $\eta = |(2e/\hbar) \sigma^{S}_{xy} / \sigma^{A}_{xy} |$.
It should be pointed out that the above decomposition of the Hall conductivity into the simple 
spin-up and spin-down contributions ($\sigma^{H\uparrow}_{xy}$ and $\sigma^{H\downarrow}_{xy}$)
is valid only for metals containing light elements such as 3$d$ transition metals.
Because the SOC generally mixes spin-up and spin-down states, 
the Hall conductivity contains both the spin-conserving part and spin-non-conserving
(spin-flipping) part\cite{Zha11}, and the occurence of the pronounced spin-non-conserving
contribution would ruin the perfect two-current model (i.e., the above simple decomposition).
The calculated spin-non-conserving contribution is indeed large in L1$_0$ FePt but becomes
small already in L1$_0$ FePd\cite{Zha11}.

The calculated $\sigma^{H\uparrow}_{xy}$, $\sigma^{H\downarrow}_{xy}$, $P^H$ and $\eta$ are listed in Table 2.
Interestingly, Table 2 indicates that unlike the longitudinal charge current, the spin-up and spin-down 
Hall currents would flow in opposite directions in all the Co-based Heusler compounds considered except Co$_2$MnIn. 
Nonetheless, the spin-up Hall conductivity is more than ten times larger than the spin-down Hall
conductivity in these compounds except Co$_2$CrGa. In other words, the Hall current carriers are mostly of
spin-up particles. Remarkably, this gives rise to the nearly 100 \%  spin polarization of the Hall current 
in all the Heusler compounds except Co$_2$CrGa (Table 2). Furthermore, the spin current to charge current ratio $\eta$
in these compounds is also very high, being just over 1.0. Therefore, both the $P^H$ and $\eta$ of the Hall current
in all the Co-based Heusler compounds except Co$_2$CrGa are close to or even better than the corresponding values
of the longitudinal current from an ideal half-metallic ferromagnet, even although some of them are not half-metallic.
This suggests that the charge Hall current from these compounds is promising for spintronic applications. 
Surprisingly, the spin-down Hall 
conductivity is non neglibile even in the half-metallic metals, corroborating that the occupied states 
well below the $E_F$ would also contribute to the anomalous Hall effect\cite{Yao04}. 
 
The spin polarization of a magnetic material is usually described in terms of the spin-decomposed DOSs at
the Fermi level as follows
\begin{equation}
P^{D} =\frac{N_{\uparrow}(E_F)-N_{\downarrow}(E_F)}{N_{\uparrow}(E_F)+N_{\downarrow}(E_F)},
\end{equation}
where $N_{\uparrow}(E_F)$ and $N_{\downarrow}(E_F)$ are the spin-up and spin-down DOSs at the $E_F$, respectively. 
The spin polarization $P^{D}$ would then vary from -1.0 to 1.0 only. For the half-metallic
materials, $P^{D}$ equals to either -1.0 or 1.0.
The calculated $N_{\uparrow}(E_F)$ and $N_{\downarrow}(E_F)$ for the Heusler compounds are listed in Table 1,
and the corresponding $P^{D}$ are listed in Table 2. In terms of $P^{D}$, only Co$_2$CrAl, Co$_2$CrSi, Co$_2$CrGe,
Co$_2$MnSi and Co$_2$MnGe are half-metallic. Interestingly, we notice that in non-half-metallic metals 
Co$_2$MnAl, Co$_2$MnGa, Co$_2$MnIn and Co$_2$MnSn, 
the $P^{D}$ is significantly smaller than the $P^{H}$, which is close to 100 \% (see Table 2).
Therefore, we may perhaps regard these compounds as anomalous Hall half metals.
As pointed out by researchers before\cite{Mazin99}, the spin polarization $P^{D}$ defined by Eq. (3) 
is not necessarily the spin polarization of the transport currents measured in the experiments. 
This can be clearly illustrated by magnetically anisotropic materials such as hcp Co. 
In hcp Co, the spin-decomposed DOSs at the $E_F$ and hence $P^{D}$ are independent 
of magnetization direction whilst, in contrast, the Hall conductivities and hence $P^{H}$
change dramatically as the magnetization rotates.\cite{Tung12}
From the viewpoint of spintronic applications, only the current spin polarizations such as $P^{H}$, instead of
the $P^{D}$, count. 

\section{Conclusions}
The anomalous and spin Hall conductivities as well as the electronic and magnetic properties of Co-based 
full Heusler compounds Co$_2$XZ (X = Cr and Mn; Z = Al, Si, Ga, Ge, In and Sn) have been calculated within the 
DFT with the GGA. The highly accurate FLAPW method is used. 
Interestingly, it is found that the spin-up and spin-down Hall currents would flow in opposite directions 
in all the Co-based Heusler compounds considered except Co$_2$MnIn, although
the spin-up Hall conductivity is more than ten times larger than the spin-down Hall
conductivity in these compounds except Co$_2$CrGa. 
Resultantly, the charge Hall current in all the Heusler compounds considered except Co$_2$CrGa 
would be almost fully spin-polarized, even although Co$_2$MnAl, Co$_2$MnGa, Co$_2$MnIn and Co$_2$MnSn are not half-metallic
from the viewpoint of the electronic structure.
Moreoever, the ratio of the accompanying spin current to the charge Hall current is slightly larger than 1.0.
Based on these results, therefore, these Heusler compounds may be called anomalous Hall half-metals.
These anomalous Hall half-metals, especially Co$_2$MnAl, Co$_2$MnGa  and Co$_2$MnIn which have large anomalous
and spin Hall conductivities, could find valuable applications in spintronics
such as magnetoresistive and spin-torque switching nanodevices as well as spin valves. 
The calculated electronic band structures and magnetic moments as well as anomalous
and spin Hall conductivities as a function of the Fermi level, are used to analyze these interesting findings.
It is hoped that these interesting results would stimulate further experimental investigations into
the anomalous Hall effect and also the characteristics of the Hall current in these Co-based Heusler compounds.

\section*{Acknowledgments}
The authors acknowledge supports from the National Science Council and the NCTS of Taiwan as well as the 
Center for Quantum Science and Engineering, National Taiwan University (CQSE-10R1004021). They also thank the
National Center for High-performance Computing of Taiwan for providing the CPU time.


\begin{thebibliography}{10}
\bibitem{Wol01}Wolf S A, Awschalom D D, Buhrman R A, Daughton J M, 
 von Molnar S, Roukes M L, Chtchelkanova and Treges 2001 {\it Science} {\bf 294} 1488
\bibitem{Gru86} Gr\"{u}berg P, Schreiber R, Pang Y, Brodsky M B and Sowers H 1986
{\it Phys. Rev. Lett.} {\bf 57} 2442
\bibitem{Bai88} Baibich M N, Broto J M, Fert A, Nguyen Van Dau F, Petroff F,
Etiemme P, Creuzet G, Friederich A and Chazelas J 1988 {\it Phys. Rev. Lett.} {\bf 61} 2472
\bibitem{Moo95} Moodera J S, Kinder L R, Wong T M and Meservy R 1995 {\it Phys. Rev. Lett.} {\bf 74} 3273
\bibitem{Parkin04} Parkin S S, Kaiser C, Panchura A, Rice P M, Hughes B, Samant M and Yang S H
 2004 {\it Nat. Matter.} {\bf 3} 862 
\bibitem{Slo96} Slonczewski J 1996 {\it J. Magn. Magn. Mater.} {\bf 159} L1
\bibitem{Mye99} Myers E B, Ralph D C, Katine J A, Louie R N, Buhrman R A 1999 {\it Science} {\bf 285} 867
\bibitem{Groot} de Groot R A, Mueller F M, van Engen P G and Buschow K H J 1983
 {\it Phys. Rev. Lett.} {\bf 50} 2024 
\bibitem{Sch86} Schwarz K 1986 {\it J. Phys. F: Met. Phys.} {\bf 16} L211
\bibitem{Brown00} Brown P J, Neumann K U, Webster P J and Ziebeck K R A 2000
 {\it J. Phys.: Condens. Matter} {\bf 12} 1827 
\bibitem{Galanakis02} Galanakis I and Dederichs P H 2002 {\it Phys. Rev.} B {\bf 66} 174429 
\bibitem{Jen03} Jeng H-T and Guo G Y 2003 {\it Phys. Rev.} B {\bf 67} 094438
\bibitem{Wan06} Wang Y K and Guo G Y 2006 {\it Phys. Rev.} B {\bf 73} 064424
\bibitem{Kubler07} K{\"u}bler K, Fecher G H and Felser C 2007 {\it Phys. Rev.} B {\bf 76} 024414
\bibitem{Kan07} Kandpal H C, Fecher G H and Felser C 2007 {\it J. Phys. D: Appl. Phys.} {\bf 40} 1507 
\bibitem{Ji01} Ji Y, Strijkers G J, Yang F Y, Chien C L, Byers J M, Anguelouch A, Xiao G, and Gupta A 
 2001 {\it Phys. Rev. Lett.} {\bf 86} 5585
\bibitem{Hal81} Hall E H 1881 {\it Philos. Mag.} {\bf 12} 157
\bibitem{Nagaosa10} Nagaosa N, Sinova J, Onoda S, MacDonald A H and Ong N P 2010
{\it Rev. Mod. Phys.} {\bf 82} 1539 
\bibitem{Yao04} Yao Y, Kleinman L, MacDonald A H, Sinova J, Jungwirth T, Wang D-S, Wang E and  Niu Q 2004
{\it Phys. Rev. Lett.} {\bf 92} 037204 
\bibitem{Rom09} Roman E, Mokrousov Y and Souza I 2009 {\it Phys. Rev. Lett.} {\bf 103} 097203
\bibitem{Fuh11} Fuh H-R and Guo G Y 2011 {\it Phys. Rev.} B {\bf 84} 144427 
\bibitem{Tung12} Tung J-C, Fuh H-R and Guo G Y 2012 {\it Phys. Rev.} B {\bf 86} 024435 
\bibitem{zen06} Zeng C, Yao Y, Niu Q and Weitering H H 2006 {\it Phys. Rev. Lett.} {\bf 96} 037204 
\bibitem{He12} He P, Ma L, Shi Z, Guo G Y, Zheng J-G, Xin Y and Zhou S M 2012 {\it Phys. Rev. Lett.} {\bf 109} 066402
\bibitem{Kubler12} K{\"u}bler J and Felser C 2012 {\it Phys. Rev.} B {\bf 85} 012405 
\bibitem{Vidal11} Vidal E V, Stryganyuk G, Schneider H, Felser C, and Jakob G 2011 {\it Appl. Phys. Lett.} {\bf 99} 132509 
\bibitem{Mur03} Murakami S, Nagaosa N, and Zhang S-C 2003 {\it Science} {\bf 301} 1348
\bibitem{Liu12} Liu L, Pai C-F, Li Y, Tseng H W, Ralph D C and Buhrman R A 2012 {\it Science} {\bf 336} 555  
\bibitem{Guo05} Guo G Y, Yao Y and Niu Q 2005 {\it Phys. Rev. Lett.} $\bf 94$ 226601 
\bibitem{Guo08} Guo G Y, Murakami S, Chen T-W and Nagaosa N 2008 {\it Phys. Rev. Lett.} {\bf 100} 096401 
\bibitem{wien2k02} Blaha P, Schwarz K, Madsen G, Kvasnicka D and Luitz J 2002  {\it WIEN2K, An Augmented Plane Wave Local Orbitals Program for Calculating Crystal Properties} (Technische University Wien, Austria, 2002)
\bibitem{Perdew96} Perdew J P, Burke K and Ernzerhof M 1996 {\it Phys. Rev. Lett.} {\bf 77} 3865 
\bibitem{Bus83} Bushow K H J and van Engen 1983 {\it J. Magn. Magn. Mater.} {\bf 38} 1 
\bibitem{Tung12b} The lattice constants of Co$_2$CrSi, Co$_2$CrGe and Co$_2$MnIn were 
determined theoretically by the GGA\cite{Perdew96} calculations
by using the accurate projector augmented-wave method, as implemented in the VASP package\cite{vasp1,vasp2}.
A large plane-wave cut-off energy of 350 eV and a 12$\times$12$\times$12 k-point mesh were used.
\bibitem{vasp1} Kresse G and Hafner J 1993 {\it Phys. Rev.} B {\bf 48} 13115 
\bibitem{vasp2} Kresse G and Furthm\"{u}ller J 1993 {\it Comput. Mater. Sci.} {\bf 6} 15 
\bibitem{Ume08} Umetsu R Y, Kobayashi K, Fujita A, Kainuma R, and Ishida K 2008
 {\it J. Appl. Phys.} {\bf 103} 07D718
\bibitem{Blochl94} Bl\"{o}chl P B, Jepseno O, and Anderson O K 1994 {\it Phys. Rev.} B {\bf 49} 16223 
\bibitem{Mav04} Mavropoulos Ph, Sato K, Zeller R, and Dederichs P H 2004 {\it Phys. Rev.} B {\bf 69} 054424
\bibitem{Kit68} Kittel C 1968 {\it Introduction to Solid State Physics} (Wiley, New York) p.470.
\bibitem{Husmann06} Husmann A and Singh L J 2006 {\it Phys. Rev.} B {\bf 73} 172417
\bibitem{kud08} Kudryavtsev Y V, Uvarov V N, Oksenenko V A, Lee Y P, Kim J B, Hyun Y H,
 Kim K W, Rhee J Y, Dubowik J 2008 {\it Phys. Rev.} B {\bf 77} 195104 
\bibitem{Wei11} Weischenberg J, Freimuth F, Sinova J, Bl{\"u}gel S, and Mokrousov Y 2011
 {\it Phys. Rev. Lett.} {\bf 107} 106601
\bibitem{See10} Seemann K M, Mokrousov Y, Aziz A, Miguel J, Kronast F, Kuch W, Blamire M G, Hindmarch A T,
 Hickey B J, Souza I and Marrows C H 2010 {\it Phys. Rev. Lett.} {\bf 104} 076402
\bibitem{Zha11} Zhang H, Freimuth F, Bl{\"u}gel S, Mokrousov Y and Souza I 2011 {\it Phys. Rev. Lett.} {\bf 106} 117202 
\bibitem{Mazin99} Mazin I I 1999 {\it Phys. Rev. Lett.} {\bf 83} 1427
\end{thebibliography}


\end{document}